\documentclass[a4paper]{article}

\usepackage{INTERSPEECH2020}
\graphicspath{{figures/}}
\usepackage{ragged2e}
\usepackage{hyperref}

\title{ Jointly Fine-Tuning ``BERT-like'' Self Supervised Models to Improve Multimodal Speech Emotion Recognition}

\name{Shamane Siriwardhana$^1$,Andrew Reis$^1$,Rivindu Weerasekera$^1$,Suranga Nanayakkara$^1$}

\address{
  $^1$ Augmented Human Lab, Auckland Bioengineering Institute, The University of Auckland}
\email{shamane@ahlab.org,andrew@ahlab.org,rivindu@ahlab.org,suranga@ahlab.org}

\begin{document}

\maketitle

\begin{abstract}
Multimodal emotion recognition from speech is an important area in affective computing. Fusing multiple data modalities and learning representations with limited amounts of labeled data is a challenging task. In this paper, we explore the use of modality specific``BERT-like'' pretrained Self Supervised Learning (SSL) architectures to represent both speech and text modalities for the task of multimodal speech emotion recognition. By conducting experiments on three publicly available datasets (IEMOCAP, CMU-MOSEI, and CMU-MOSI), we show that jointly fine-tuning ``BERT-like'' SSL architectures achieve state-of-the-art (SOTA) results. We also evaluate two methods of fusing speech and text modalities and show that a simple fusion mechanism can outperform more complex ones when using SSL models that have similar architectural properties to BERT.
\end{abstract}

\section{Introduction}
Emotion recognition plays a significant role in many intelligent interfaces~\cite{picard2000affective}. Even with the recent advances in Deep Learning (DL), this is still a challenging task. The main reason being that most publicly available annotated datasets in this domain are small in scale, which makes DL models prone to over-fitting. Another important feature of emotion recognition is the inherent multi-modality in the way we express emotions~\cite{li2019attentive}. Emotional information can be captured by studying many modalities, including facial expressions, body postures, and EEG~\cite{sebe2005multimodal}. Of these, arguably, speech is the most accessible. In addition to accessibility, speech signals contain many other emotional cues~\cite{kim2018emotion}. Although speech signals contain substantial amounts of information, it can be unrewarding to drop the linguistic component that coexists with it, especially given that the text component can be easily transcribed in real world applications with the considerable successes in the domain of speech-to-text with several commercial-scale APIs being available~\cite{singh2019asroil}.

In multimodal emotion recognition, representation learning and fusion of modalities can be identified as a major research area~\cite{tsai2018learning,zadeh2018multi,tsai2019multimodal}. Recent work has explored the use of deep representations in contrast to low level representations~\cite{swain2018databases} such as MFCC, COVAREP~\cite{degottex2014covarep} or GloVe embeddings~\cite{pennington2014glove}. Such deep representation techniques can mainly be categorised into two main categories: 1) transfer learning techniques that use pretrained networks to extract features~\cite{han2019transfer,parthasarathy2019improving,feng2020review} or fine-tune models~\cite{lu2019speech}; and 2) unsupervised embeddings learning techniques which include variational auto-encoders (VAE)~\cite{latif2017variational} and adversarial auto-encoders (AE)~\cite{sahu2018adversarial}. It is also important to highlight that performance usually degrades in transfer learning techniques due to the mismatch of source and target tasks. Recent work~\cite{locatello2018challenging} explains the problems related to learn disentangled representations from VAE when no inductive bias on the model or the dataset exists. In terms of fusing multiple modalities, recent work has explored architectures like attention~\cite{li2019attentive,zadeh2018multi}, graph neural networks~\cite{ghosal2019dialoguegcn} and transformers~\cite{tsai2019multimodal}.
Multimodal fusion mechanisms, especially those that fuse deep representations, usually result in architecturally complex models~\cite{zhang2019multimodal}.

In representation learning, a class of techniques known as SSL has achieved SOTA performance in many areas of Natual Language Processing (NLP)~\cite{devlin2018bert,liu2019roberta}, Computer Vision~\cite{kolesnikov2019revisiting,lu2019vilbert} (CV) and speech recognition~\cite{pascual2019learning,schneider2019wav2vec,baevski2019vq}. SSL enables us to use a large unlabelled dataset to train models that can be later used to extract representations and fine-tune for specific problems that may have limited amounts of training data. Prior works~\cite{devlin2018bert,kolesnikov2019revisiting} have highlighted the effectiveness of fine-tuning pre-trained SSL models for specific tasks in contrast to using them only as frozen feature extractors.
A significant transition happened in the field of NLP with the introduction of SSL models like the Deep Bidirectional Transformers~\cite{devlin2018bert} (BERT) and its successors~\cite{liu2019roberta}. By adding a single task-specific layer to a pre-trained SSL model like BERT, one can solve multiple downstream tasks. BERT-like models also consist of favourable architectural features such as the $CLS$ token, which can be used as a representation for the entire sequence. Another important factor is the extensive availability of pre-trained models in the open-source community, which leads to both cost and time savings since these models tend to be very computationally expensive to train from scratch.

Even though several SSL models have been introduced for speech recognition related tasks like speech-to-text~\cite{schneider2019wav2vec,baevski2019vq} and speech emotion recognition~\cite{pascual2019learning}, prior work has not looked at combining multiple separate SSL models, each specializing in one modality. This may be due to the architectural complexity of such models brought on by the fact that usually, these SSL networks have a large number of parameters. Combining multiple drastically different high dimensional representations is also not a simple task and may increase the parameter count even further. If, however, the modality specific SSL architectures share similar properties, then we may be able to use simpler fusion mechanisms to extract information for the desired task.  Our work was heavily inspired by recent work~\cite{baevski2019vq,baevski2019effectiveness}, which explored the effectiveness of self supervised pre-training with discretized speech representations for the task of Automatic Speech Recognition (ASR).

For the first time in the literature, we jointly fine-tuned modality-specific ``BERT-like'' SSL models that represent speech~\cite{baevski2019effectiveness,baevski2019vq} and text~\cite{lu2019vilbert} on the task of multimodal emotion recognition. We further evaluate how simple fusion methods, -
which add minimal additional trainable parameters, performed when compared with more complex fusion mechanisms such as Co-Attentional~\cite{lu2019vilbert} fusion. We also conducted a series of ablation studies to explore which factors affect the performance of these models. Please refer the \href{https://github.com/shamanez/BERT-like-is-All-You-Need}{\textit{Pytorch implementation.}}

\section{Pre-Trained SSL models}
We summarise the three pretrained SSL models that were used to process speech and text signals in the proposed framework in the following sections. We use pretrained models available in the Fairseq toolkit~\cite{ott2019fairseq}.


\subsection{VQ-Wav2Vec}
VQ-Wav2Vec~\cite{baevski2019vq} is an extension of Wav2Vec~\cite{schneider2019wav2vec}, which focuses on moving continuous speech representations into the discrete domain. Wav2Vec~\cite{schneider2019wav2vec} learns representations from speech signals based on Contrastive Predictive Coding~\cite{oord2018representation} (CPC). The major difference in VQ-Wav2Vec~\cite{baevski2019vq} from Wav2Vec~\cite{schneider2019wav2vec} is the application of Vector Quantization~\cite{van2017neural} methods to generate discretized speech representations. In our experiments, we used a pretrained VQ-Wav2Vec~\cite{baevski2019vq} model that trained on Librispeech-960~\cite{panayotov2015librispeech} to represent speech signals as a sequence of tokens, similar to the tokenization step for a sentence in NLP.

\subsection{Speech-BERT}

The term \textit{Speech-BERT} is used in our work to define a BERT-like Transformer architecture trained on a set of discretized speech tokens, where the speech signal was discretized and tokenized by a pretrained VQ-Wav2Vec as mentioned in the above section. We were heavily motivated by recent work\cite{baevski2019effectiveness}, which illustrated the effectiveness of BERT-like models in the domain of ASR. We used a pretrained Speech-BERT, which was trained on the discretized Librispeech-960~\cite{panayotov2015librispeech} dataset with the pretext task of mask token prediction. The Speech-BERT model architecture is similar to BERT-base~\cite{devlin2018bert} that consists of 12 layers and an embedding dimension of 768. During our experiments, we fine-tune the Speech-BERT model for the task of multimodal emotion recognition.

\subsection{RoBERTa model}

 RoBERTa~\cite{liu2019roberta} is an extension of the BERT~\cite{devlin2018bert} model which does not use the next sentence prediction task~\cite{liu2019roberta} during training.
 The RoBERTa~\cite{liu2019roberta} architecture consists of 24 layers and an embedding dimension of 1024. Similar to Speech-BERT, we fine-tune the RoBERTA~\cite{liu2019roberta} model for the task of multimodal emotion recognition.


\section{Methodology}

We explore the use of Speech-BERT and RoBERTa SSL models for the task of multimodal speech emotion recognition. As the first step, we evaluate two possible fusion mechanisms to combine the two SSL models. The performance of the final proposed model was then compared with published SOTA results on IEMOCAP~\cite{busso2008iemocap}, CMU-MOSEI~\cite{zadeh2018multimodal}, and CMU-MOSI~\cite{zadeh2016mosi} datasets. Finally, we conduct an extensive set of ablation studies with the IEMOCAP dataset~\cite{busso2008iemocap} to understand the behaviour of our proposed framework under different settings. We investigate the effects of fine-tuning and frozen states for Speech-BERT, RoBERTa, as well as the effects of different fusion mechanisms.

\subsection{Model Pipeline}

\begin{figure}[!t]
\centering
\includegraphics[width=0.35\textwidth]{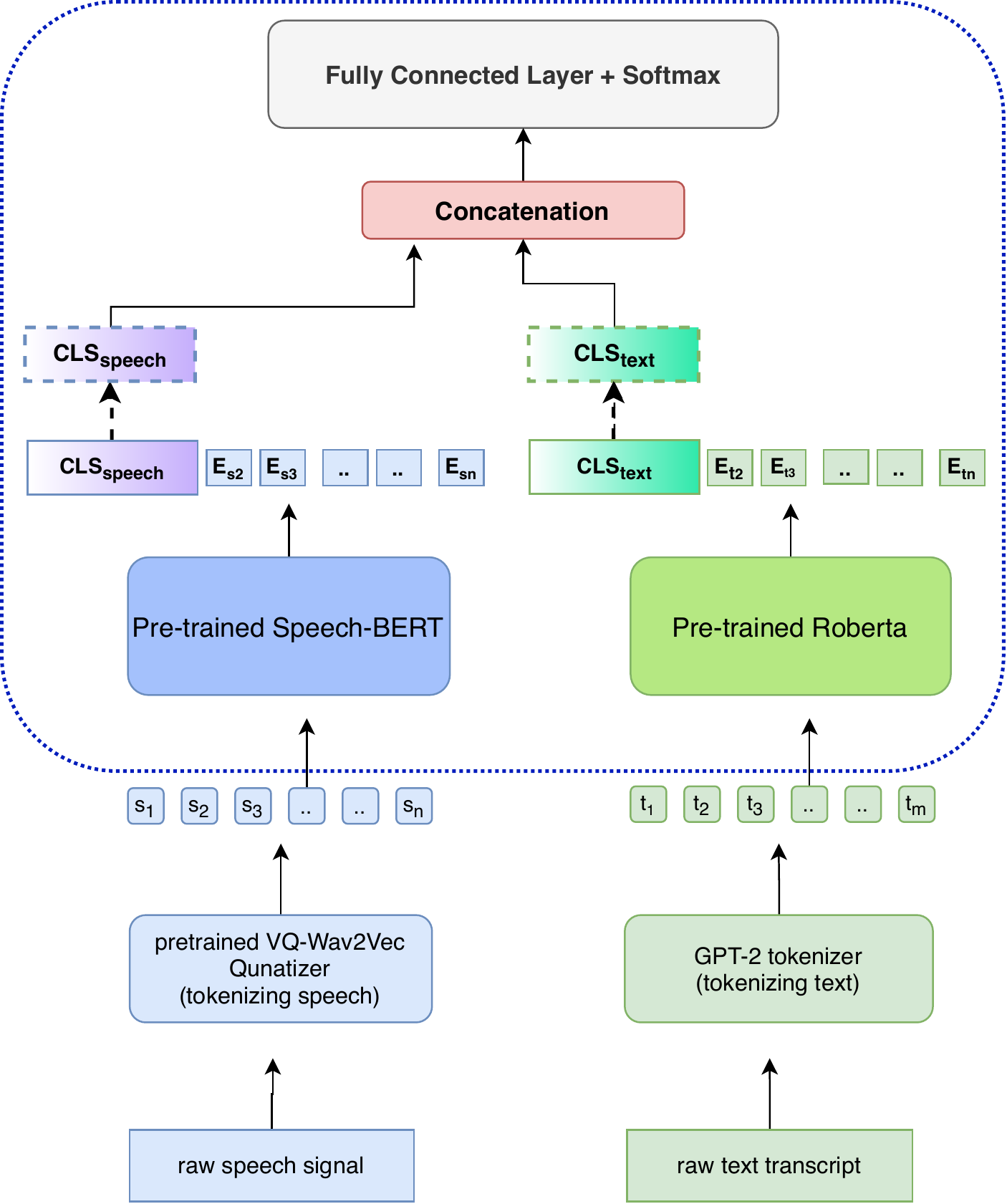}
\caption{Overview of the proposed framework with Shallow-Fusion}
 \vspace{-10pt}
\justify
\label{fig:overview}
\end{figure}

Figure~\ref{fig:overview} gives an overview of the proposed framework. Both the speech signal and text transcripts are simultaneously fed into the model via two different pipelines. The speech signal gets discretized by a pre-trained VQ-Wav2Vec~\cite{baevski2019vq} model and the text transcript is tokenized with the GPT-2 tokenizer~\cite{radford2019language}. Once speech and text modalities are tokenized, we send them through pre-trained Speech-BERT and RoBERTa models, where the outputs have embedding sizes 768 and 1024 and maximum sequence lengths of 2048 and 512 respectively. The next step is fusing these embeddings prior to the prediction head, which consists of a single fully connected layer. In this paper we explore two possible mechanisms, which we will discuss in Section~\ref{sec:fuse}. Finally, we fine-tune the entire framework, including both Speech-BERT and Roberta SSL models (the components inside the blue dotted box in Figure~\ref{fig:overview}).

\subsection{Fusion of SSL model outputs}\label{sec:fuse}

The fusion mechanism plays an essential role in any multimodal speech emotion recognition framework. In this work, we analyse how the following different fusion mechanisms affect performance.

\subsubsection{Shallow Fusion}

The success of BERT in sentence classification tasks~\cite{devlin2018bert} highlights the effective use of the $CLS$ token as a representation of the entire sentence. $CLS$ stands for the classification and it is the first token of every input sequence to the BERT. Motivated by recent work in the domain of NLP~\cite{devlin2018bert,liu2019roberta}, we concatenate the two $CLS$ tokens computed respectively from speech-BERT and RoBERTa models as described in Figure~\ref{fig:overview} . Finally we send the concatenated embedding through a classification head that includes a fully connected layer that outputs logits followed by a softmax function. Due to the simplicity of the fusion, we describe this mechanism as Shallow-Fusion. We also use Shallow-Fusion as the standard fusion mechanism in the ablation studies since it achieved superior performances in our experiments.

\subsubsection{Co-Attentional Fusion}

\begin{figure}[t!]
\centering
\includegraphics[width=0.46\textwidth]{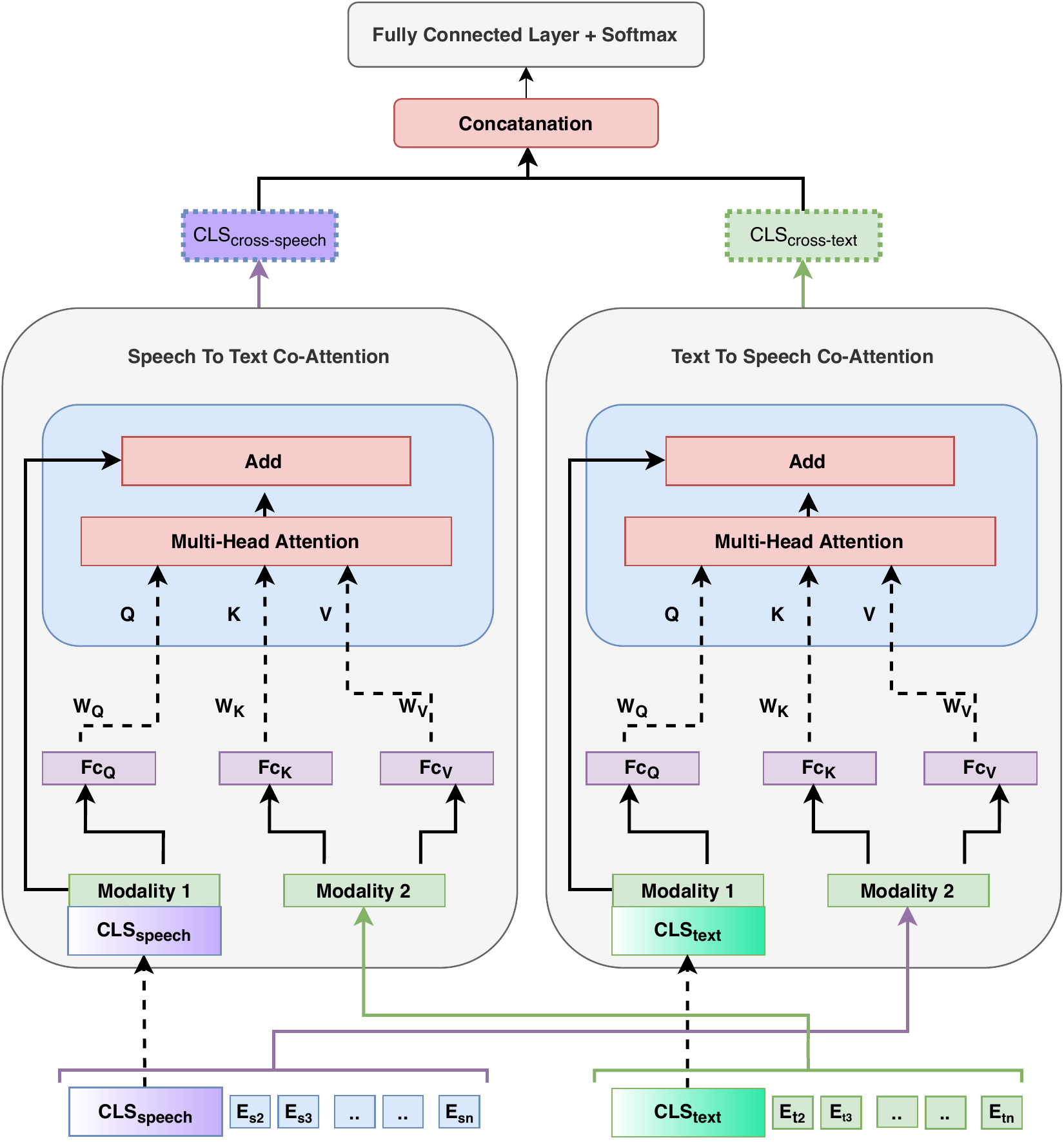}
\caption{Co-Attentional layer and fusion mechanism}
\label{fig:co-fusion}
\vspace{-10pt}
\end{figure}

In order to provide an opportunity for an embedding level interaction between the two modalities, we propose to use a Co-Attentional layer~\cite{lu2019vilbert}. Co-Attention is a variant of Self-Attention~\cite{vaswani2017attention} that has been used in visual-linguistic Transformers like VilBERT~\cite{lu2019vilbert}. In contrast to Self-Attention, Co-Attention is computed by interchanging Key-Value vector pairs from one modality with the Query vector from another modality. Since the $CLS$ token from each modality already aggregates the sequential information~\cite{devlin2018bert}, we calculate the Query vector from each mortality's $CLS$ token and let it attend to the entire sequence of the other modality embeddings. After the Co-Attentional layer, we concatenate the modified $CLS$ tokens from each modality and send these through a prediction head. Figure~\ref{fig:co-fusion} gives a detailed illustration of the Co-Attentional layer and the fusion mechanism.


%
%
%
\section{Experimental Setup}

We implemented our model using Pytorch and the Fairseq~\cite{ott2019fairseq} toolkit. All models were trained under distributed settings, using two Tesla v100 32GB GPUs with an effective batch-size of 16.
The Adam optimiser was used in the optimization with warm-up updates and the polynomial decay learning-rate scheduler. The initial learning rate and dropout were set to $1e^{-5}$ and $0.1$.

\subsection{Performance Comparison with SOTA}

\subsubsection{IEMOCAP Experiments}

The IEMOCAP~\cite{busso2008iemocap} dataset contains conversation data of 10
male and female actors. Similar to prior work~\cite{poria2017context,li2019attentive}, we selected the most commonly used four emotion categories of Happy (\& Excitement), Sad, Anger, and Neutral.
We followed the experimental procedure and evaluation metrics of previous studies~\cite{tsai2019multimodal,sun2019learning}.
Table~\ref{tab:IEMOCAP} provides a comparison of model performance with other SOTA models on Binary Accuracy (BA) and the F1-score~\cite{sun2019learning,tsai2019multimodal}.  Table~\ref{tab:IEMOCAP-4 class} illustrates the performance comparison w.r.t 4-class unweighted accuracy metric following the recent work done by Li et al~\cite{li2019attentive}.

\subsubsection{CMU-MOSEI and CMU-MOSI Experiments}

CMU-MOSEI~\cite{zadeh2018multimodal} is the current largest dataset for multimodal emotion recognition that consists of $22,000$ examples created by extracting review videos from YouTube. Each example is annotated with an integer score between -3 to +3. CMU-MOSI~\cite{zadeh2016mosi} also has similar properties to CMU-MOSEI~\cite{zadeh2018multimodal} but only with 2000 examples. To compare our model with both datasets, we follow the latest prior work that has used these datasets~\cite{tsai2019multimodal,sun2019learning}. For both, we used 7-class accuracy, Mean-Average-Error (MAE), 2-class accuracy (binary), and F1-score. Table~\ref{tab:MOSEI} and Table~\ref{tab:MOSI} describe the evaluation results on CMU-MOSEI~\cite{zadeh2018multimodal} and CMU-MOSI~\cite{zadeh2016mosi} datasets respectively.


\section{Ablation studies}

We conducted four ablation studies using the IEMOCAP~\cite{busso2008iemocap} dataset to understand the behaviour of the proposed framework. We use Binary-Accuracy and F1 score for each emotion as the evaluation metric. In order to have a fair setting in our ablation experiments, we use the first three sessions of the IEMOCAP~\cite{busso2008iemocap} dataset as training, the fourth session as validation and fifth session as the test set. Table~\ref{tab:ABLATION} illustrates the results of our four ablation studies discussed in the following sections.

\subsection{Comparison between two fusion mechanisms in Fine-Tune state}
\vspace{-5pt}

In the first ablation study, Table~\ref{tab:ABLATION} (5.1), we compare the performance of each fusion mechanism when fine-tuning Speech-Bert and Roberta for the downstream task of multimodal-emotion recognition. Shallow-Fusion shows a slight improvement with respect to binary accuracy and F1-scores for each emotion over Co-Attentional fusion. This illustrates how a simple fusion mechanism followed by a classification head work remarkably well with finetuned ``Bert-Like"" pretrained SSL models even in a multimodal setting.


\subsection{Comparison between Unimodal inputs}
\vspace{-5pt}

\begin{figure*}[!t]
\begin{minipage}[h]{1.0\textwidth}
\centering
\includegraphics[width=\textwidth]{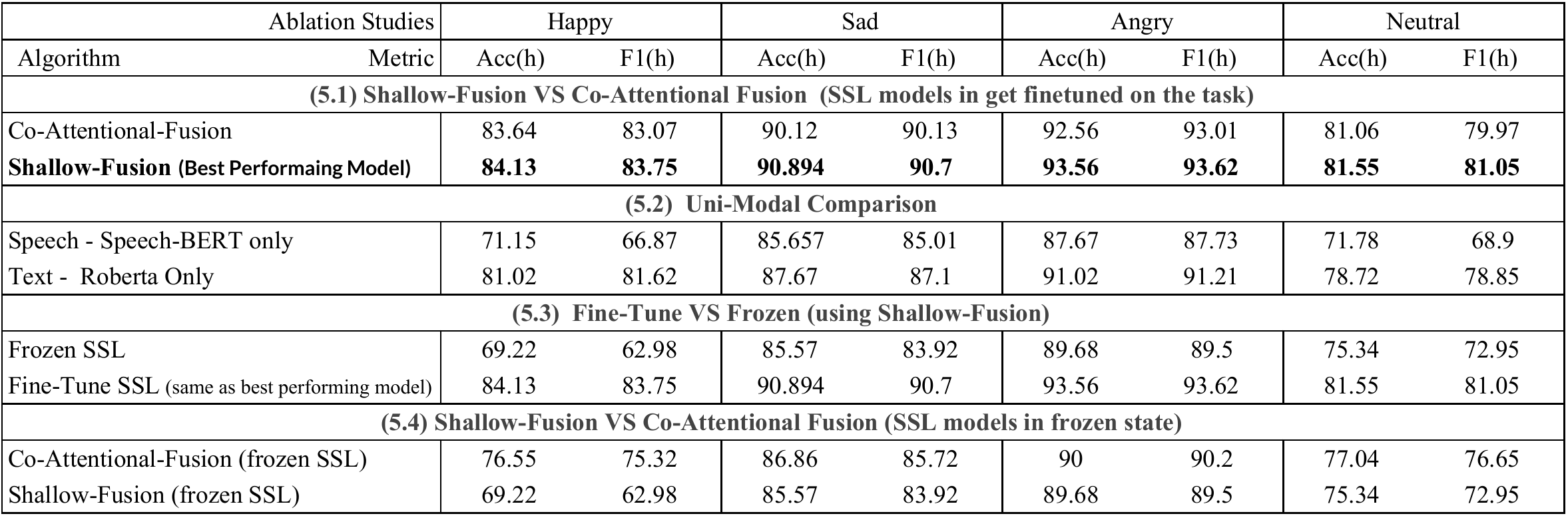}
\captionof{table}{Evaluation results of the ablation studies on IEMOCAP dataset with Binary Accuracy (BA) and F1 score }
\vspace{-10pt}
\label{tab:ABLATION}
\centering
\end{minipage}

\vspace{1.00mm}

\begin{minipage}[h]{1.0\textwidth}
\centering
\includegraphics[width=\textwidth]{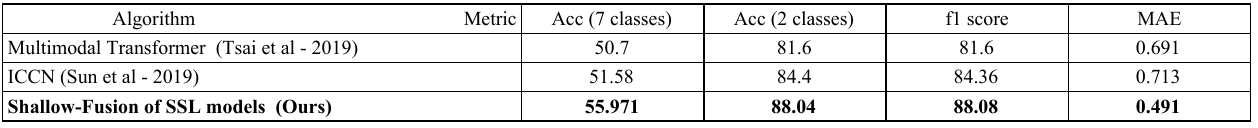}
\captionof{table}{ Results for multimodal emotions analysis on CMU-MOSEI with Seven Class accuracy ,BA (binary accuracy) F1 score, and MAE. Performance's of the other models are taken from ICCCN (2019)~\cite{sun2019learning} MULT (2019)~\cite{tsai2019multimodal}}
\vspace{-10pt}
\centering
\label{tab:MOSEI}
\end{minipage}

\vspace{1.00mm}

\begin{minipage}[h]{1.0\textwidth}
\centering
\includegraphics[width=\textwidth]{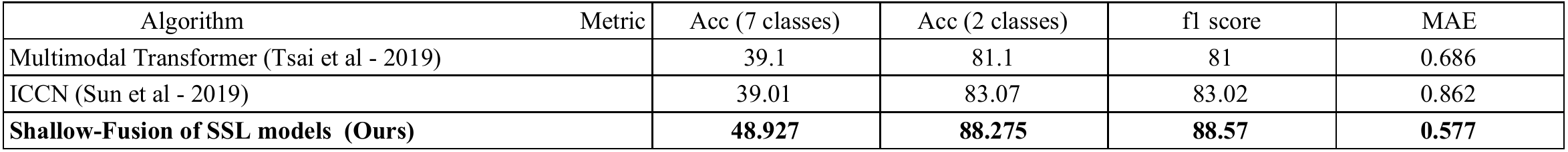}
\captionof{table}{ Results for multimodal emotions analysis on CMU-MOSI. Performance of the other models are taken from ICCCN (2019)~\cite{sun2019learning} MULT (2019)~\cite{tsai2019multimodal}}
\vspace{-10pt}
\centering
\label{tab:MOSI}
\end{minipage}

\vspace{1.00mm}

\begin{minipage}[h]{1.0\textwidth}
\centering
\includegraphics[width=\textwidth]{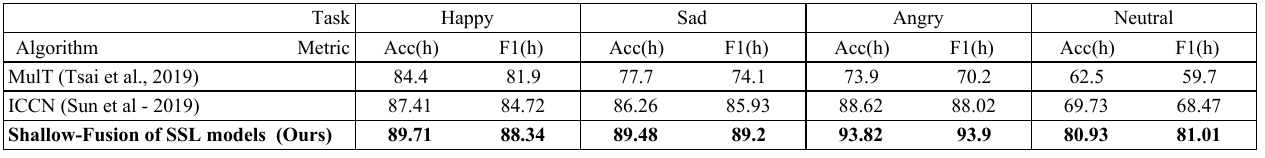}
\captionof{table}{ Results for multimodal emotions analysis on IEMOCAP with Binary Accuracy (BA) ad F1 score. Performance of the other models are taken from ICCCN (2019)~\cite{sun2019learning} MULT (2019)~\cite{tsai2019multimodal}}
\vspace{-10pt}
\label{tab:IEMOCAP}
\centering
\end{minipage}

\vspace{1.00mm}

\begin{minipage}[h]{1.0\textwidth}
\centering
\includegraphics[width=\textwidth]{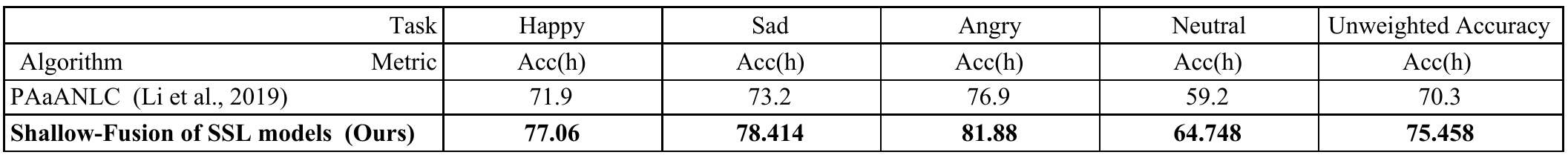}
\captionof{table}{ Results for multimodal emotions analysis on IEMOCAP with 4 class unweighted accuracy. Performance of the other model is taken from PAaAN (2019)~\cite{li2019attentive}}
\vspace{-10pt}
\label{tab:IEMOCAP-4 class}
\centering
\end{minipage}
\end{figure*}

Table~\ref{tab:ABLATION} (5.2) illustrates the performance comparison between uni-modal inputs. In this experiment, we use the $CLS$ token of Speech-BERT and Roberta as the sequence representation for speech and text. As the results suggest, text-only performs better than speech-only. A possible reason might be the availability of high emotional clues in the linguistic structure. However, we can still see a clear improvement when comparing text-only results with our best performing multimodal model, which highlights the importance of multi-modality.

\subsection{Comparison between fine-tuning and frozen SSL architectures}

\vspace{-5pt}
Table~\ref{tab:ABLATION}  (5.3) looks at the effect of finetuning vs having frozen Speech-BERT and RoBERTa models. Unsurprisingly, fine-tuning the two SSL models with Shallow-Fusion for the downstream task leads to better performance.

\subsection{Comparison between the two fusion mechanisms when keeping SSL models frozen}
\vspace{-5pt}

Finally, we compare how the two fusion mechanism behave when we keep Speech-BERT and RoBERTa in a frozen state -using them only as feature extractors. Table~\ref{tab:ABLATION} (5.4) shows when the SSL networks are frozen, Co-Attentional fusion performs better.
We highlight that the increased number of interactions prior to the prediction layers enables Co-Attentional fusion to adapt better than Shallow-Fusion.
Co-Attentioal fusion requires a much larger number of new trainable parameters. While the Co-Attentional layer adds nearly 6 million new parameters, Shallow-Fusion only adds close to fourteen thousand parameters.

\vspace{-10pt}
\section{Conclusion}
In this work, we use two pretrained ``BERT-like" architectures to solve the downstream task of multimodal emotion recognition. As per our knowledge, this is the first time that two SSL algorithms that represent speech and texts are fine-tuned for the task of multimodal speech emotion recognition. By conducting several experiments, we show how a simple fusion mechanism (Shallow-Fusion) makes the overall framework simple and straightforward and improve on more complex fusion mechanisms. We also highlight the importance of introducing BERT-like models to process speech signals, which can easily be used to improve the performance of multimodal tasks like emotion recognition. Having structurally similar ``BERT-like" architectures to represent both speech and text allows us to fuse modalities in a straightforward way and quickly adapt standard practices in the NLP domain.

In future work, we hope to visualize and further explore the behavior of SSL models for the task of multimodal emotion recognition. Exploring the use of BERT-like models to represent speech could enable advances in NLP to be easily used in the domain of speech.

\section{Acknowledgements}
This work is supported by the Assistive Augmentation research grant under the Entrepreneurial Universities (EU) initiative of New Zealand.

\newcommand{\BIBdecl}{\setlength{\itemsep}{0.01 em}}
\bibliographystyle{IEEEtran}
{\footnotesize\bibliography{mybib}}

\end{document}